\documentclass[twoside,twocolumn,9pt]{article}
\usepackage{extsizes}
\usepackage[super,sort&compress,comma]{natbib} 
\usepackage[version=3]{mhchem}
\usepackage[left=1.5cm, right=1.5cm, top=1.785cm, bottom=2.0cm]{geometry}
\usepackage{balance}
\usepackage{mathptmx}
\usepackage{sectsty}
\usepackage{graphicx} 
\usepackage{lastpage}
\usepackage[format=plain,justification=justified,singlelinecheck=false,font={stretch=1.125,small,sf},labelfont=bf,labelsep=space]{caption}
\usepackage{float}
\usepackage{fancyhdr}
\usepackage{fnpos}
\usepackage[english]{babel}
\addto{\captionsenglish}{%
  
}
\usepackage{array}
\usepackage{droidsans}
\usepackage{charter}
\usepackage[T1]{fontenc}
\usepackage[usenames,dvipsnames]{xcolor}
\usepackage{setspace}
\usepackage[compact]{titlesec}
\usepackage{hyperref}

\usepackage{epstopdf}

\definecolor{cream}{RGB}{222,217,201}

\begin{document}

\pagestyle{fancy}
\thispagestyle{plain}
\fancypagestyle{plain}{
\renewcommand{\headrulewidth}{0pt}
}

\makeFNbottom
\makeatletter
\renewcommand\LARGE{\@setfontsize\LARGE{15pt}{17}}
\renewcommand\Large{\@setfontsize\Large{12pt}{14}}
\renewcommand\large{\@setfontsize\large{10pt}{12}}
\renewcommand\footnotesize{\@setfontsize\footnotesize{7pt}{10}}
\makeatother

\renewcommand{\thefootnote}{\fnsymbol{footnote}}
\renewcommand\footnoterule{\vspace*{1pt}%
\color{cream}\hrule width 3.5in height 0.4pt \color{black}\vspace*{5pt}} 
\setcounter{secnumdepth}{5}

\makeatletter 
\renewcommand\@biblabel[1]{#1}            
\renewcommand\@makefntext[1]%
{\noindent\makebox[0pt][r]{\@thefnmark\,}#1}
\makeatother 
\renewcommand{\figurename}{\small{Fig.}~}
\sectionfont{\sffamily\Large}
\subsectionfont{\normalsize}
\subsubsectionfont{\bf}
\setstretch{1.125} 
\setlength{\skip\footins}{0.8cm}
\setlength{\footnotesep}{0.25cm}
\setlength{\jot}{10pt}
\titlespacing*{\section}{0pt}{4pt}{4pt}
\titlespacing*{\subsection}{0pt}{15pt}{1pt}

\fancyfoot{}
\fancyfoot[LO,RE]{\vspace{-7.1pt}\includegraphics[height=9pt]{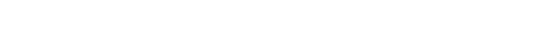}}
\fancyfoot[CO]{\vspace{-7.1pt}\hspace{13.2cm}\includegraphics{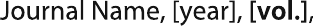}}
\fancyfoot[CE]{\vspace{-7.2pt}\hspace{-14.2cm}\includegraphics{head_foot/RF}}
\fancyfoot[RO]{\footnotesize{\sffamily{1--\pageref{LastPage} ~\textbar  \hspace{2pt}\thepage}}}
\fancyfoot[LE]{\footnotesize{\sffamily{\thepage~\textbar\hspace{3.45cm} 1--\pageref{LastPage}}}}
\fancyhead{}
\renewcommand{\headrulewidth}{0pt} 
\renewcommand{\footrulewidth}{0pt}
\setlength{\arrayrulewidth}{1pt}
\setlength{\columnsep}{6.5mm}
\setlength\bibsep{1pt}

\makeatletter 
\newlength{\figrulesep} 
\setlength{\figrulesep}{0.5\textfloatsep} 

\newcommand{\topfigrule}{\vspace*{-1pt}%
\noindent{\color{cream}\rule[-\figrulesep]{\columnwidth}{1.5pt}} }

\newcommand{\botfigrule}{\vspace*{-2pt}%
\noindent{\color{cream}\rule[\figrulesep]{\columnwidth}{1.5pt}} }

\newcommand{\dblfigrule}{\vspace*{-1pt}%
\noindent{\color{cream}\rule[-\figrulesep]{\textwidth}{1.5pt}} }

\makeatother

\twocolumn[
  \begin{@twocolumnfalse}
{\includegraphics[height=30pt]{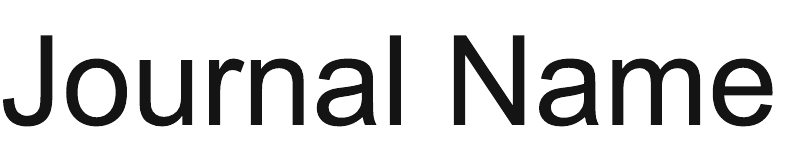}\hfill\raisebox{0pt}[0pt][0pt]{\includegraphics[height=55pt]{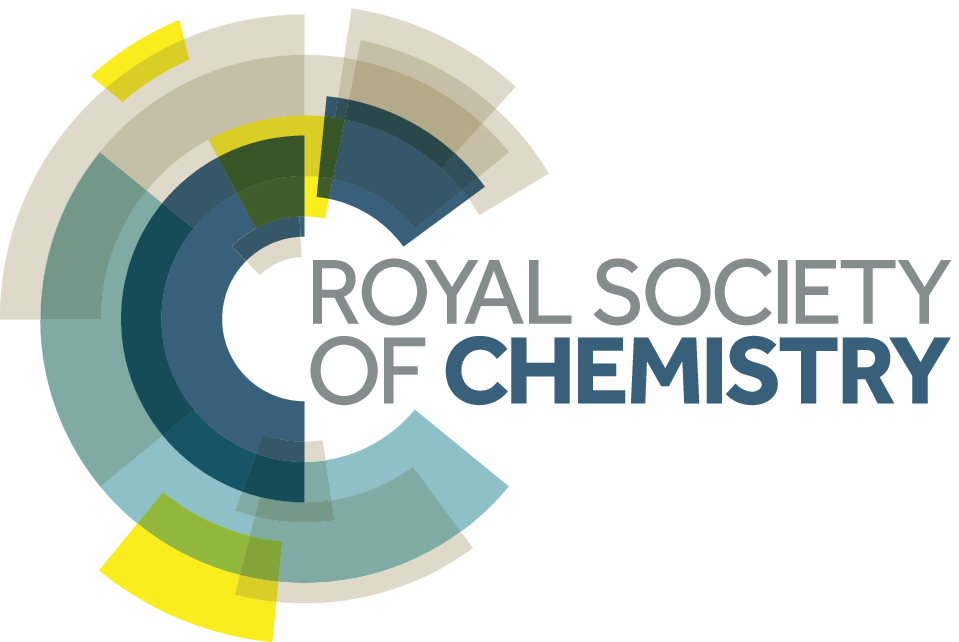}}\\[1ex]
\includegraphics[width=18.5cm]{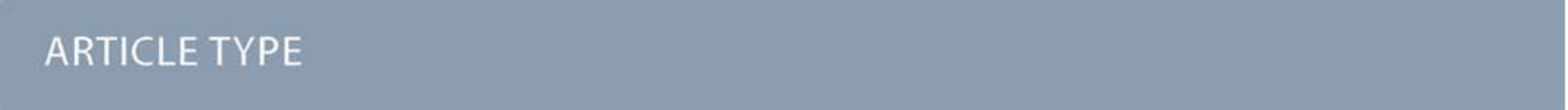}}\par
\vspace{1em}
\sffamily
\begin{tabular}{m{4.5cm} p{13.5cm} }

\includegraphics{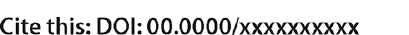} & \noindent\LARGE{\textbf{Dynamic Kerr and Pockels Electro-Optics of Liquid Crystals in Nanopores for Active Photonic Metamaterials$^\dag$}} \\
\vspace{0.3cm} & \vspace{0.3cm} \\

 & \noindent\large{Andriy~V.~Kityk,$^{\ast}$\textit{$^{a}$} Marcjan~Nowak,\textit{$^{a}$} Manuela Reben,\textit{$^{b}$} Piotr~Pawlik,\textit{$^{c}$} Monika~Lelonek,\textit{$^{d}$} Anatoliy~Andrushchak,\textit{$^{e}$} Yaroslav~Shchur,\textit{$^{f}$} Nazariy~Andrushchak,\textit{$^{g}$} and Patrick~Huber$^{\ast}$\textit{$^{hij}$}} \\

\includegraphics{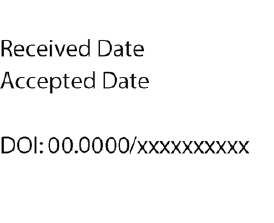} & \noindent\normalsize{Photonic metamaterials with properties unattainable in base materials are already beginning to revolutionize optical component design. However, their exceptional characteristics are often static, as artificially engineered into the material during the fabrication process. This limits their application for in-operando adjustable optical devices and active optics in general.
Here, for a hybrid material consisting of a liquid crystal-infused nanoporous solid, we demonstrate active and dynamic control of its meta-optics by applying alternating electric fields parallel to the long axes of its cylindrical pores. First-harmonic Pockels and second-harmonic Kerr birefringence responses, strongly depending on the excitation frequency- and temperature, are observed in a frequency range from 50~Hz to 50~kHz. This peculiar behavior is quantitatively traced by a Landau-De Gennes free energy analysis to an order-disorder orientational transition of the rod-like mesogens and intimately related changes in the molecular mobilities and polar anchoring at the solid walls on the single-pore, meta-atomic scale. Thus, our study evidences that liquid crystal-infused nanopores exhibit integrated multi-physical couplings and reversible phase changes that make them particularly promising for the design of photonic metamaterials with thermo-electrically tunable birefringence in the emerging field of spacetime metamaterials aiming at a full spatio-temporal control of light.} 
\end{tabular}
\end{@twocolumnfalse} \vspace{0.6cm}

  ]

\renewcommand*\rmdefault{bch}\normalfont\upshape
\rmfamily
\section*{}
\vspace{-1cm}


\footnotetext{\textit{$^{a}$~Faculty of Electrical Engineering, Czestochowa
University of Technology, Al. Armii Krajowej 17, 42-200 Czestochowa, Poland, E-mail: andriy.kityk@univie.ac.at.}}
\footnotetext{\textit{$^{b}$~Faculty of Materials Science and Ceramics, AGH-University of Science and Technology, al. Mickiewicza 30, 30-059 Cracow, Poland. }}
\footnotetext{\textit{$^{c}$~Faculty of Production Engineering and Materials Science, Czestochowa
University of Technology, Al. Armii Krajowej 19, 42-200 Czestochowa, Poland. }}
\footnotetext{\textit{$^{d}$~SmartMembranes GmbH, Heinrich-Damerow-Str. 4, 06120 Halle(Saale), Germany. }}
\footnotetext{\textit{$^{e}$~Institute of Telecommunications, Radioelectronics and Electronic Engineering, Lviv Polytechnic National University, 12 Bandery St., Lviv 79046, Ukraine. }}
\footnotetext{\textit{$^{f}$~Institute for Condensed Matter Physics, 1 Svientsitskii str., 79011 Lviv, Ukraine. }}
\footnotetext{\textit{$^{g}$~SPC SoftPartners, 97 Konovalca str., 79057 Lviv, Ukraine. }}
\footnotetext{\textit{$^{h}$~Hamburg University of Technology, Institute for Materials and X-Ray Physics, 21073 Hamburg, Germany, E-mail:patrick.huber@tuhh.de. }}
\footnotetext{\textit{$^{i}$~Deutsches Elektronen-Synchrotron DESY, Centre for X-Ray and Nano Science CXNS, 22607 Hamburg, Germany. }}
\footnotetext{\textit{$^{j}$~Hamburg University, Centre for Hybrid Nanostructures CHyN,  22607 Hamburg, Germany. }}




\section{Introduction}
2-D and 3-D metamaterials offer exciting opportunities for the manipulation of light propagation and light-matter interactions \cite{Yu2014, Kadic2019, Caloz2020}. Although metamaterials have proven to be a versatile platform for passive optics, active controlling of their optical properties is still a challenging task. 

Here phase-change materials, in particular employing inorganic, hard matter with an amorphous to crystalline transition open up many opportunities for implementing externally tunable metamaterials as the reversible transition is accompanied by a considerable change of the refractive index \cite{Lepeshov2021TunableMetasurfaces, Zhang2021, Wang2021}. Also embedding liquid crystals (LCs) in nanoporous solids allows one to combine self-assembly and phase transition behavior of organic, soft matter with a scaffold structure that provides mechanical robustness, and thus to fabricate soft-hard hybrids with tunable multi-physical couplings in terms of self-organization, mechanical, electrical and optical properties \cite{Ruiz-Clavijo2021RevisitingApplications., Ryu2017, Sentker2019Self-AssemblyMetamaterials, Spengler2018, Xu2021}. 

If the pore size is below the visible light wavelength the single pores act as meta-atoms. The effective optics is then determined by the pore shape, size and orientation. Moreover, the pore filling allows one to adjust the light-matter interaction on a meta-atomic scale \cite{Sautter2015, Sentker2019Self-AssemblyMetamaterials, Waszkowska2021, Xu2021}. Effective optical properties not achievable in base materials are possible \cite{Liu2010Self-AlignmentApplications,Yu2011LightRefraction, Zheludev2012,Sautter2015, Jalas2017,Guo2017Nanoscale, Kadic2019,Lininger2020OpticalCrystals,Maccaferri2020, Xu2021}, specifically adjustable birefringence and thus polarization-dependent light propagation speeds \cite{Sentker2019Self-AssemblyMetamaterials}. 
However, the resulting optics is also here often static as dictated by the fabrication process and can not easily be dynamically adapted, as it is demanded for the emerging field of active and spacetime photonic metamaterials aimed at a full spatio-temporal control of light \cite{Shaltout2019, Caloz2020, Engheta2021MetamaterialsMore,Wang2021}. Here, we follow the strategy to overcome this static behavior by applying external electrical fields to liquid crystals confined in anodic aluminum oxide nanopores.

\begin{figure}[htbp]
\centering
\includegraphics[width=0.95\columnwidth]{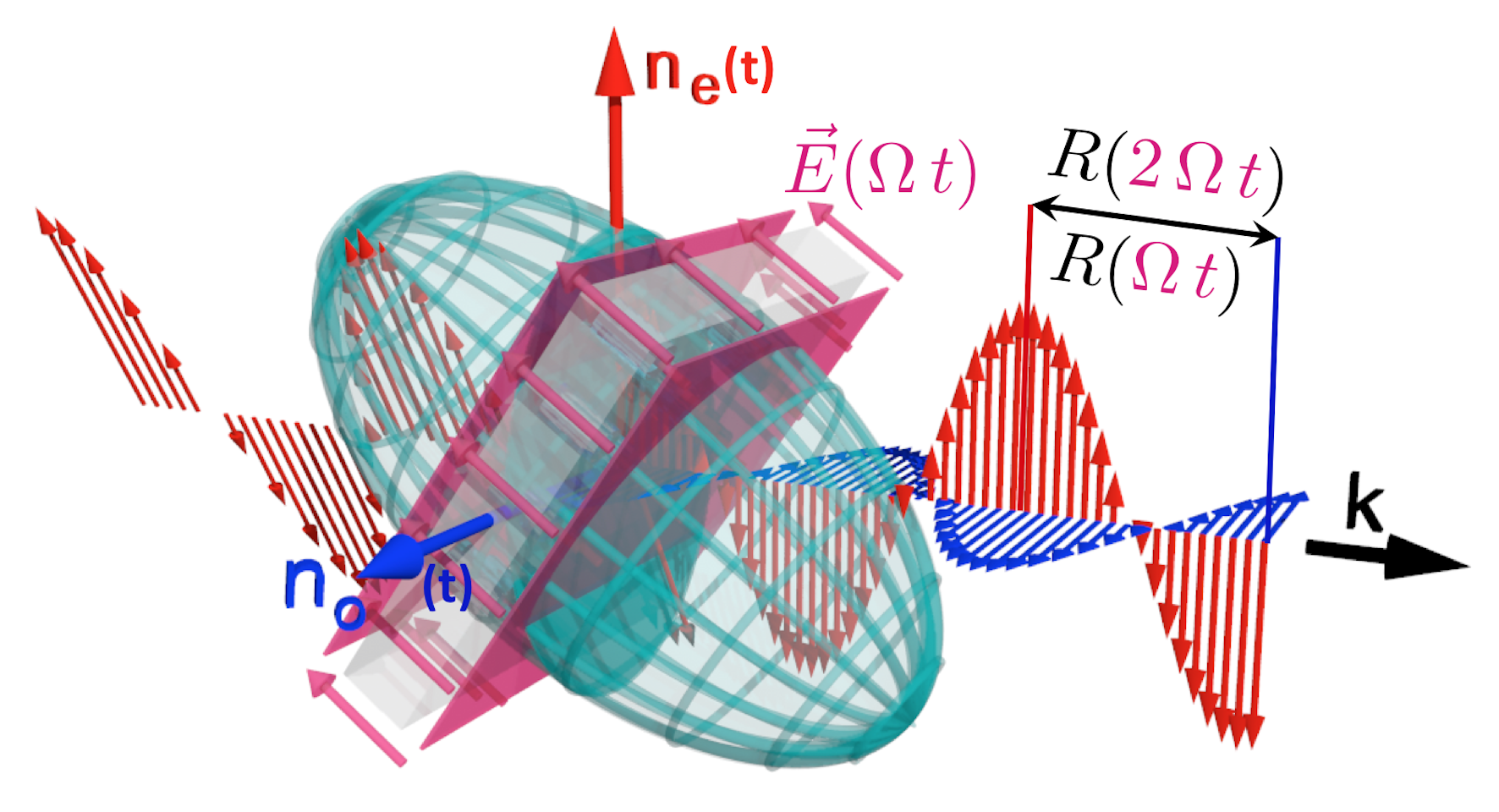}
\caption{\textbf{Dynamic electro-optics of a liquid crystal-infused nanoporous photonic metamaterial.} (a) A linearly polarized laser beam is split up into two waves, i.e., the extraordinary (red) and ordinary wave (blue), when it traverses a material consisting of liquid crystal-infused cylindrical nanopores. The optical anisotropy, as illustrated by the ellipsoid of refractive indices (indicatrix), results in different propagation speeds and thus in a phase difference (retardation) $R$ between the two transmitted waves after passing the birefringent medium. Here the extraordinary beam is slower than the ordinary beam resulting in a positive phase retardation $R$ of the electromagnetic waves after passing the pore array. This corresponds to a prolate indicatrix and a ''positive" optical birefringence, i.e., $\Delta n=n_{\rm e}-n_{\rm o}>0$. If the liquid-crystalline filling is exposed to an external electrical field $E(\Omega)$ alternating with a circular frequency $\Omega$, it responses with molecular reorientations and the phase shift of the transmitted wave components consists of two frequencies, a first-harmonic, $R(1\Omega)$ and a second-harmonic response, $R(2\Omega)$ corresponding to a linear, Pockels and a quadratic, Kerr electro-optic effect, respectively. Note that for clarity only the electrical fields of the electromagnetic waves are shown.  
} \label{fig:DynElectroOptics_3DMetMat}
\end{figure}

Hereby, we have to consider that in many previous experimental and theoretical studies  \cite{Crawford1996x,Kutnjak2003,Kityk2008,Calus2012,Calus2014, Kopitzke2000,Cerclier2012,Kityk2014,Sentker2018,Yildirim2019CollectiveConfinement,Chahine2010,Grigoriadis2011, Kityk2010x,Busch2017, Wittmann2021Particle-resolvedConfinement} it has been shown that the self-assembly behavior and molecular mobility can be substantially modified in spatially confined geometries. More importantly, the impacts of these confinement effects on the electro-optics have been barely explored so far. Therefore, part of our study will also explore the phase behavior of the confined liquid crystal and its impact on the thermo-optical and electro-optical couplings in nanoconfinement.

In the following we provide a brief introduction into electro-optic effects and present then our experimental study aimed at integrating these functionalities in a nanoporous composite. The electro-optic effect was first reported by John Kerr in 1875 \cite{Kerr1875}. While examining optically isotropic media he noticed electric field-induced optical birefringence (anisotropy), the magnitude of which increased quadratically with the electric field strength. This behavior is known nowadays as Kerr's law. Eight years later, R\"ontgen and Kundt independently reported electro-optic effects in quartz and tourmaline, i.e., in anisotropic piezoelectric crystalline materials \cite{Narasimhamurti1981}. In contrast to the Kerr effect, the electrically induced birefringence changed linearly with the applied electric field. In a number of later experiments it was demonstrated that the linear electro-optic effect is observable only in non-centrosymmetric crystals and it was properly described by Pockels in 1893 within a phenomenological theory based on crystal symmetry and tensor analysis \cite{Narasimhamurti1981}. 

Thus, in crystalline materials both linear or quadratic electro-optic effects are essentially related to anisotropic effective optical properties. The corresponding tensor analysis shows that inversion symmetry implies that a linear response is forbidden. For this reason the Pockels effect is in particular observable in piezoelectric crystals with broken inversion symmetry. Moreover, it may be found in poled glasses or ceramics \cite{Yamaoka2015} and likewise in other noncrystalline polar materials as well as chiral LC materials, for example in tilted chiral smectics (ferroelectric LCs) \cite{Skarabot1998a,Skarabot1998,Skarabot1999,Busch2017}.

The Kerr effect, in contrast, is observable also in centrosymmetric media, including isotropic liquids. It is particularly large in liquids with polar molecules, such as e.g. nitrobenzene, being widely applied in electro-optical light shutters and modulators \cite{Knudsen1975}. 

Both linear Pockels and quadratic Kerr effects are caused by the rotational diffusion of the molecules, the dipoles of which are able to follow to the applied electric field. Liquids consisting of small polar molecules are characterized by rather fast electro-optic response times, usually in the range of nanoseconds or even below \cite{Righini1993,Halbout1982}, whereas the dynamics slows down for large molecules \cite{Bermudez2000,Niziol2010}.

From the application point of view both electro-optic effects provide means for a refractive index modulation driven by an applied electric field. This offers an elegant possibility to integrate the modulation of phase retardation or light intensity into a material. They are thus the base for a large variety of applications in science and technology aimed at the spatio-temporal control of light \cite{Narasimhamurti1981}. 

Here we report a dynamic electro-optical study (electro-optical time response) on the rod-like nematic LC 7CB embedded into cylindrical nanopores of anodic alumina oxide (AAO) membranes. Linear and quadratic electro-optical effects are explored via first and second harmonic response, respectively, measured by a lock-in technique in a frequency window from 50 Hz to 50 kHz and a broad temperature range encompassing paranematic and confined nematic states. These electro-optic measurements are complemented by simultaneous polarimetric measurements of the optical retardation (birefringence) exploring the temperature evolution of the static orientational order in the region of the paranematic-to-nematic transition. The obtained results are analyzed within a phenomenological Landau-De Gennes model accounting for the interaction of the nematic order parameter and the applied electric field.

\section{Experimental}
\subsection{Materials and sample preparation}
\begin{figure}[htbp]
\centering
\includegraphics[width=0.9\columnwidth]{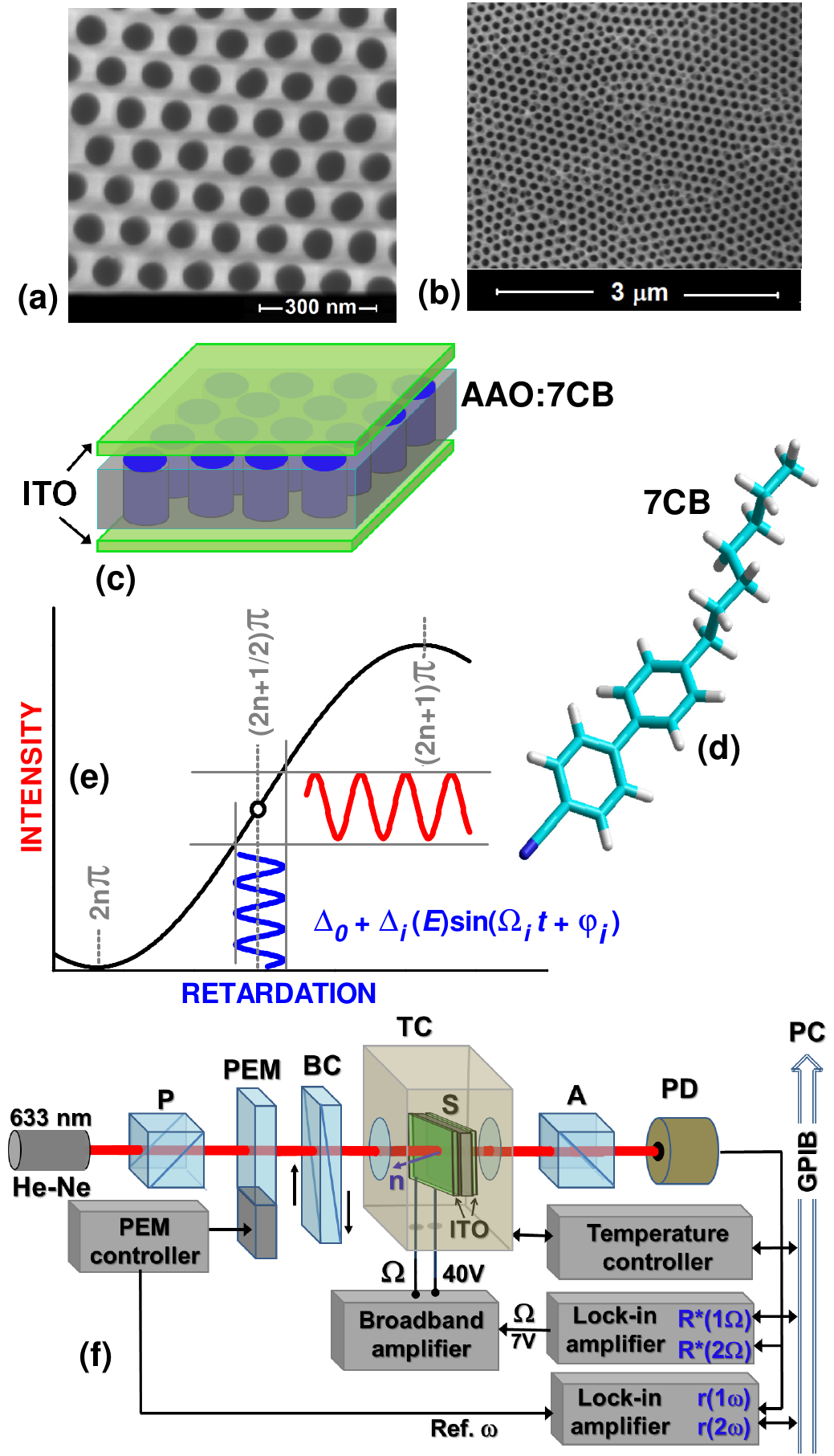}
\caption{\textbf{Electro-optic experiment and detection scheme.} (a, b) Scanning electron micrographs of the AAO membrane ($d$=65 nm) used as host matrix. (c) The electro-optic sample cell consists of an AAO membrane, filled with (d) the calamatic LC 7CB, stacked between two ITO-glass electrodes. (f) The setup to measure simultaneously the optical retardation and the dynamic electro-optic response combines a photo-elastic modulation polarimeter and a lock-in amplifier for the detection of the first (1$\Omega$) and second (2$\Omega$) harmonics corresponding to the linear (Pockels) and quadratic (Kerr) electro-optic effects, respectively. It consists of a He-Ne - helium-neon laser (633 nm), PEM - photo-elastic modulator, P - polarizer, A - analyzer, BC - Babinet compensator, TC - controlled temperature cell, S - sample (AAO:7CB membrane
stacked between ITO-glass electrodes), PD - broad band diode photodetector. (e) Dependence of the light intensity vs. the phase retardation: the modulated retardation converted to the modulated light intensity. An electro-optic sine-wave-retardation,
$\Delta_i(E)\sin(\Omega_it+\varphi_i)$ ($\Omega_i =\Omega$, 2$\Omega_2$) is converted in an undistorted way into a modulated intensity, provided that the static retardation $\Delta_0=(2n\pm1/2)\pi$  $(n=0,1,2,..)$ and $\Delta_i(E) \ll \pi$. } 
\label{fig:exp_illustration}
\end{figure}

The nematogen liquid crystal 7CB (4-heptyl-4-cyano biphenyl, see Fig.~\ref{fig:exp_illustration}) was purchased from Merck AG. A characteristic feature of 7CB is a simple phase diagram in the bulk state. It exhibits only a single first order phase transition during cooling from the isotropic (I) to the nematic (N) phase at $T_{\rm IN}$ = 315.5 K \cite{Chirtoc2004} before solidification. The anodic aluminum oxide (membrane thickness $h$ =130 $\mu$m) was synthesized by means of a double-stage electrochemical etching process of pure aluminum in an oxalic acid (30\%) electrolyte. The resulting mesoporous AAO matrices consist of a hexagonally arranged, parallel array of slightly conical channels (mean pore diameter 65~nm), see Fig.~\ref{fig:exp_illustration}. The conicity of the pores results from the unidirectional etching process \mbox{\cite{Cencha2019NondestructiveImbibition}}.

The nematic 7CB melt was embedded into the cylindrical nanopores by capillarity-driven spontaneous imbibition \cite{Gruener2011}. To build an electro-optical cell for the dynamic electro-optic and the polarimetric measurements the AAO:7CB composite membrane was clamped between two transparent ITO glass electrodes, see Fig.~\ref{fig:exp_illustration}. The electro-optic cell was placed into a temperature cell operated by a temperature controller (Lakeshore-340) with
a control accuracy of 0.01~K.

\subsection{Electro-optic setup}
In Fig.~\ref{fig:exp_illustration} the laboratory setup for simultaneous optical retardation and electro-optic measurements is shown. It combines a photo-elastic modulation polarimeter and lock-in amplifier technique for a reference registration of the first (1$\Omega$) and second (2$\Omega$) harmonics of the electro-optical response. Since the AAO membrane consists of an array of parallel tubular nanopores both the empty AAO matrices as well as the LC nanocomposites exhibit effectively an uniaxial optical anisotropy with one optical axis parallel to the long pore axis \cite{Kityk2008,Calus2014,Calus2012}. This optical uniaxiality remains unchanged both as a function of temperature or under an applied electric field. Thus the observation of changes in the optical indicatrix is only possible in a tilted membrane geometry, see Fig.~\ref{fig:DynElectroOptics_3DMetMat}. Accordingly, the sample was tilted with respect to the normal incidence of the laser beam by $\sim$ 40 degree about the incident laser beam direction, see Fig.~\ref{fig:DynElectroOptics_3DMetMat} and \ref{fig:exp_illustration}.

The measuring process switched between two detection modes, the polarimetry and the electro-optic response mode, respectively. The polarimetry mode is active when
the photo-elastic modulator (PEM, $f=$~50~kHz, PEM-90 Hind Instruments) was turned on. The optical retardation, $\Delta$, is defined as  $\Delta n\,L$, where $\Delta n$ is
the effective birefringence in the direction of light propagation and $L$ is the length of the pathway of light traveling through the anisotropic medium. In this mode a simultaneous
reference detection of the first ($r_{1f}$) and second ($r_{2f}$) harmonics of modulated light (hereafter termed PEM harmonics) by means of the lock-in amplifier (Stanford Research, SR-830) yielded the value of the optical retardation $\Delta=\arctan(kr_{1f}/r_{2f})$. Here the factor $k$, which is defined by a ratio of Bessel functions $J_2(A_0)/J_1(A_0)$ at the retardation PEM amplitude ($A_0=$~0.383~$\lambda$) and the photodetector (PD) frequency response characteristic, was determined within a calibration procedure. Similar polarimetry setups have been described in earlier studies, for additional details see Refs. \cite{Skarabot1998,Kityk2008,Busch2017x,Calus2016ChiralNanochannels, Sentker2019Self-AssemblyMetamaterials,Sentker2019Dissertation}.

\subsection{Dynamic electro-optic analysis}
In the electro-optic response mode the PEM was turned off. An alternating electric field $E_0\sin{(\Omega t)}$ ($E_0\sim$3000 V/cm) was applied parallel to the nanopore axis by applying electric potentials between the transparent ITO electrodes. This induces an electro-optic retardation. In the low frequency limit ($\Omega \tau \ll 1$) the linear electro-optic effect implies that the electro-induced retardation $\Delta_e \propto E_0\sin(\Omega t)$, i.e., its phase modulation is of the same frequency as the exciting electrical field, $\Omega$. For the quadratic
electro-optic effect, in contrast, $\Delta_e \propto E^2_0\sin^2(\Omega t)=E^2_0(1-\cos(2\Omega t))/2=E^2_0/2+(E^2_0/2)\sin(2\Omega t+\pi/2)$ the electric field-induced retardation is characterized by the static component ($\propto E^2_0$) and a dynamic component of doubled frequency, $2\Omega$ that is phase shifted by $\pi/2$ and $\propto E^2_0$. At higher frequencies relaxation processes emerge ($\Omega \tau \sim 1$) leading to a phase shift $\varphi$ in the electro-optic response.
In the polarizer-analyzer system the electric field-induced dynamic retardation is converted into a modulated light intensity, which is detected by the photodetector PD and
subsequently analyzed by the lock-in amplifier SR-830 with respect to the applied reference voltage extracting thus the amplitude ($\rho_{_{\Omega_i}}$) and phase
($\varphi_i$) of its first ($1\Omega$) and second ($2\Omega$) harmonic components, of the complex electro-optic responses,
 I.e. $R^*_{1\Omega}=\rho_{_{1\Omega}}\exp(i\varphi_1)$ and $R^*_{2\Omega}=\rho_{_{2\Omega}}\exp(i(\varphi_2+\pi/2))=i\rho_{_{2\Omega}}\exp(i\varphi_2)$, respectively.

Please note that a retardation-to-intensity conversion function in the system with crossed polarizer P and analyzer A is defined as
$\propto I_0\sin^2(\Delta/2)=I_0(1-\cos\Delta)/2$, see sketch in Fig.~\ref{fig:exp_illustration}, where $I_0$ is the incident light intensity of the coherent He-Ne laser light
($\lambda=$633~nm). The total retardation $\Delta$, in general, represents a superposition of the static retardation, $\Delta_0$, induced by the sample (S)
and the Babinet compensator (BC), as well as the dynamic electro-optic component, $\Delta_i(E_0)\sin(\Omega_it+\varphi_i)$. Accordingly, the dynamic retardation
is converted to the modulated light intensity in an undistorted way (linear conversion), if the static retardation $\Delta_0=(2n\pm1/2)\pi$  $(n=0,1,2,..)$ whereas
$\Delta_i(E_0) \ll \pi$. Taking into account that the sample birefringence changes with temperature the BC serves here to adjust precisely the static phase
shift to the condition mentioned above. The corresponding manual adjustment, which corresponds to the condition that the first PEM harmonic $r_{1f}$ reaches a 
maximum (i.e. $r_{2f}=0$), was made precisely in the polarimetry mode at each temperature set point prior to the electro-optic response measurements.

The magnitude of the first PEM harmonic, $r_{1f}(A_0=0.383\lambda)$, which is proportional to the incident light intensity $I_0$, has also been used as reference value for the normalization of the electro-optic response amplitudes $\rho_{_{1\Omega}}$ and $\rho_{_{2\Omega}}$. The electro-optic retardation amplitude $\Delta_i(E_0)$, on the other hand, did not exceed in our experiments one angular minute. Thus both requirements for linear
retardation-to-intensity conversion indeed have been perfectly fulfilled.
Note that the amplitudes $\rho_{_{1\Omega}}$ and $\rho_{_{2\Omega}}$ are normalized by a value of
$10^{-6}r_{1\omega}(0.383\lambda)$ as reference magnitude, that is extracted in polarimetry mode for a PEM retardation amplitude set to $0.383\lambda$.
 
\section{Results and discussion}
\subsection{Temperature-dependent static collective orientational order and dynamic electro-optics}
\begin{figure}[htbp]
\vspace{-0.0cm}
\centering
\includegraphics[width=0.95\columnwidth]{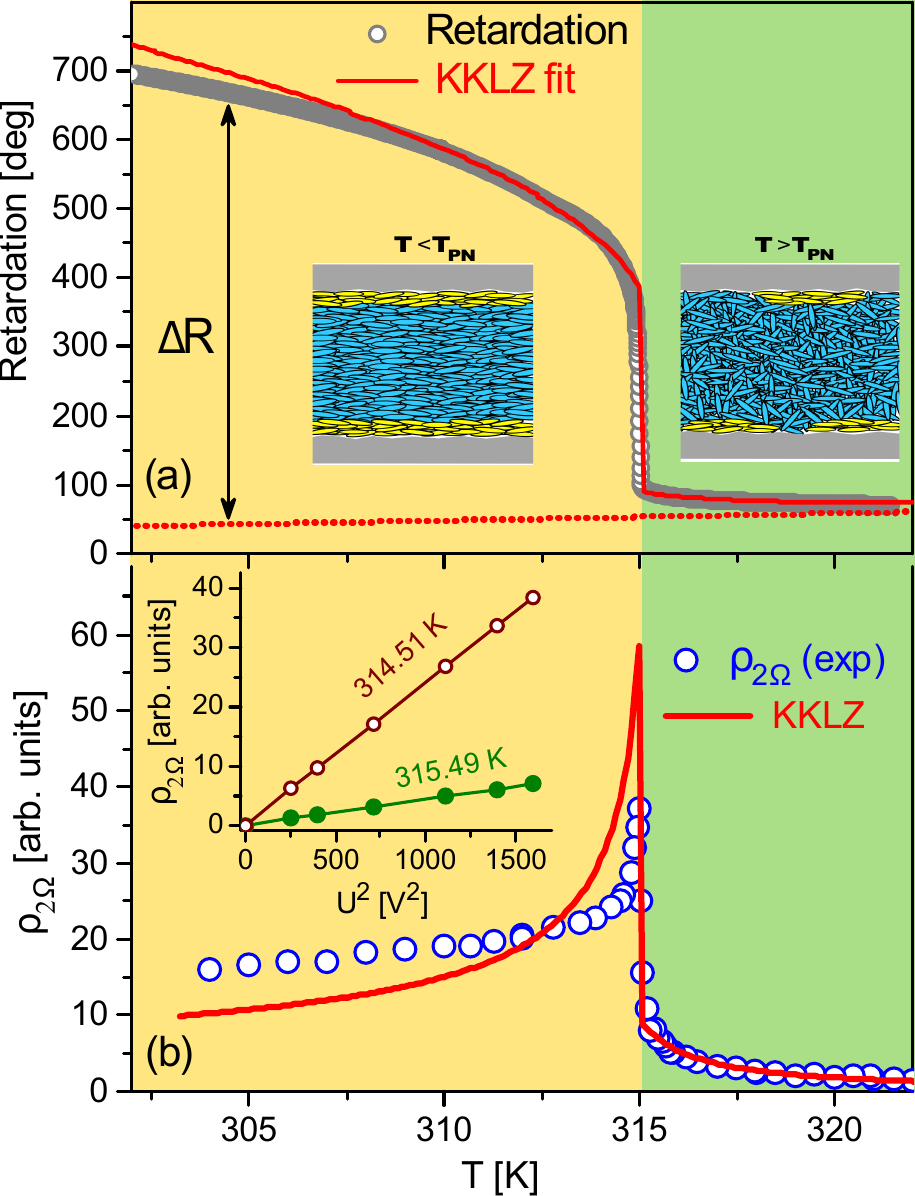}
\vspace{-0.0cm}
\caption{\textbf{Temperature-dependent static nematic order in comparison with dynamic electro-optics and the corresponding free-energy modeling in an AAO:7CB photonic metamaterial.} (a) Temperature dependence of the measured optical retardation and its theoretical description by a free energy model (KKLZ-model). (b) Temperature dependence of the measured quadratic electro-optic response (Kerr effect) and its best fit based with the KKLZ model. Inset in (b) voltage dependence of the second-harmonic electro-optic response at a temperature above and below the paranametic-to-nematic phase transition, $T_{\rm PN}$, respectively. The yellow and green backgrounds indicate the nematic and paranematic temperature ranges, respectively.}
\label{fig:KKLZ_modeling}\end{figure}

The optical retardation, $R(T)$ characterizes the evolution of the static orientational ordering inside the nanochannels ($R\propto S$, where $S$ is the nematic order parameter) as a function of temperature, see Fig.~\ref{fig:KKLZ_modeling}. In the paranematic phase ($T > T_{PN}$) $R(T)$ is small indicating an almost isotropic confined LC. However, the $R(T)$-dependence exhibits a characteristic pretransitional tail typical of nematic precursor ordering in the interfacial region of tubular nanochannels \cite{Kityk2008}. A nearly step-like increase at $T_{PN}$ indicates a discontinuous evolution of the paranematic-to-nematic ordering caused by an abrupt appearance of nematic molecular arrangement in the core region of the pore filling, see Fig.~\ref{fig:KKLZ_modeling} \cite{Kityk2008}.

\begin{figure*}[htbp]
\vspace{-0.0cm}
\centering
\includegraphics[width=0.95\columnwidth]{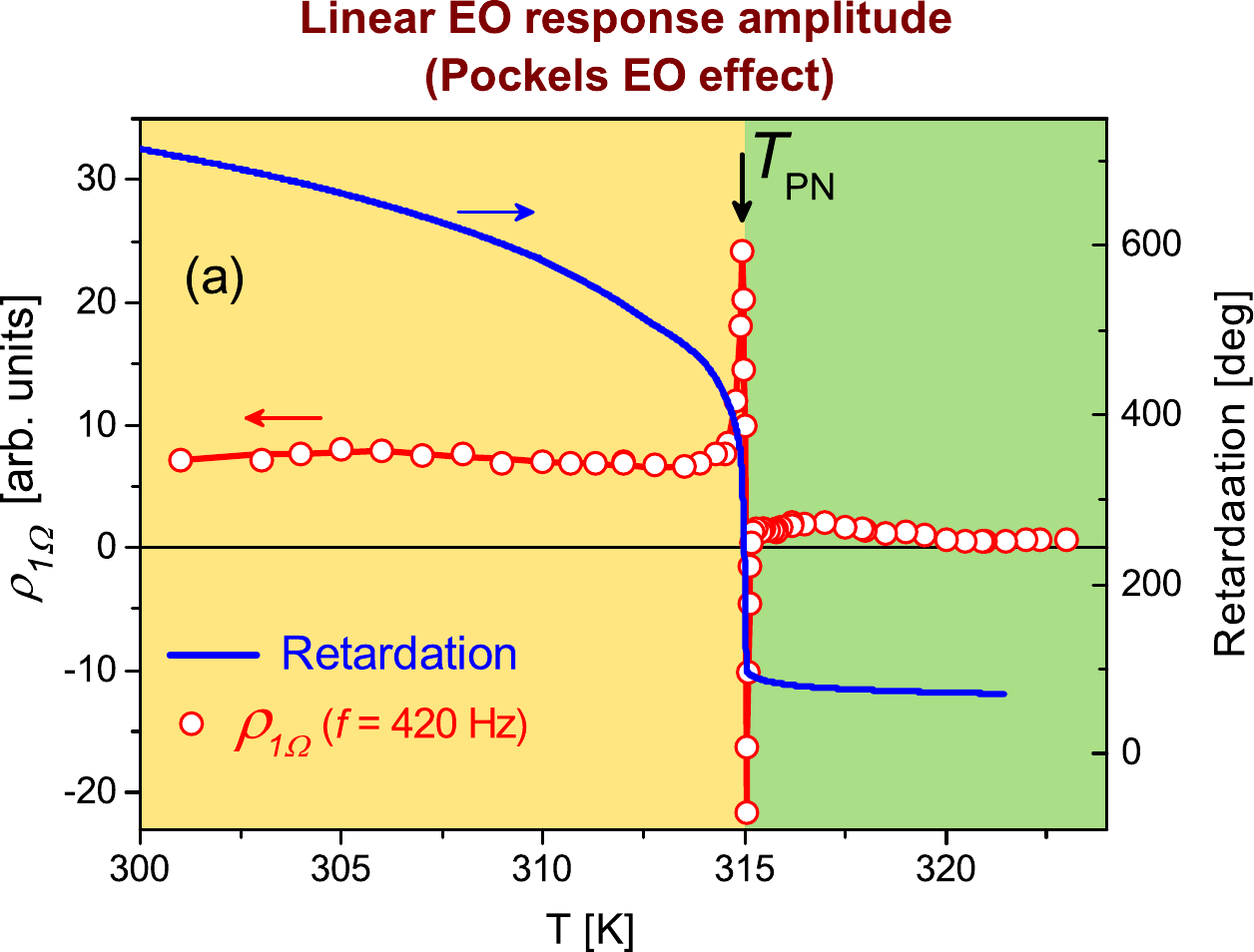}\hspace{5mm}
\includegraphics[width=0.95\columnwidth]{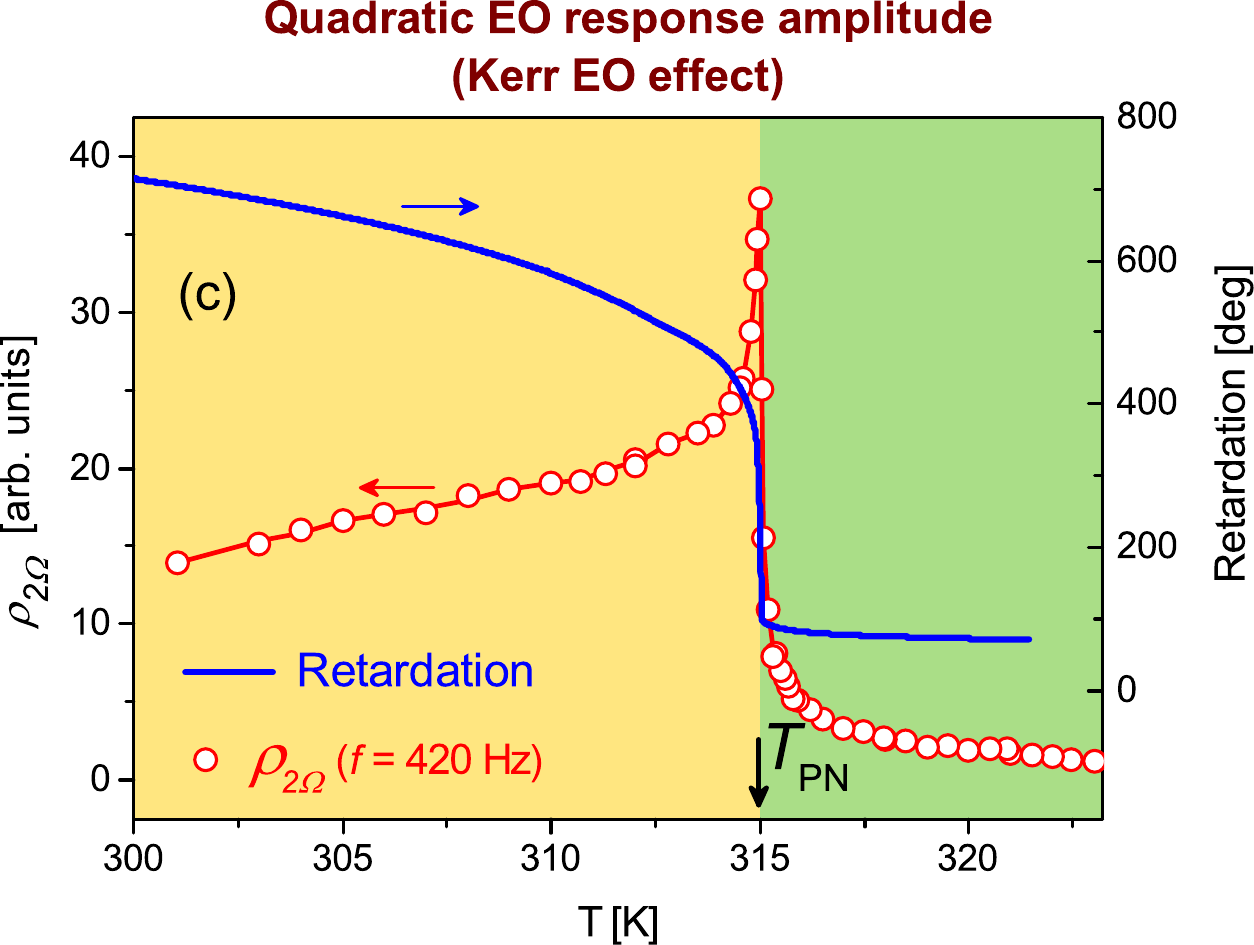}\\
\vspace{5mm}
\includegraphics[width=0.95\columnwidth]{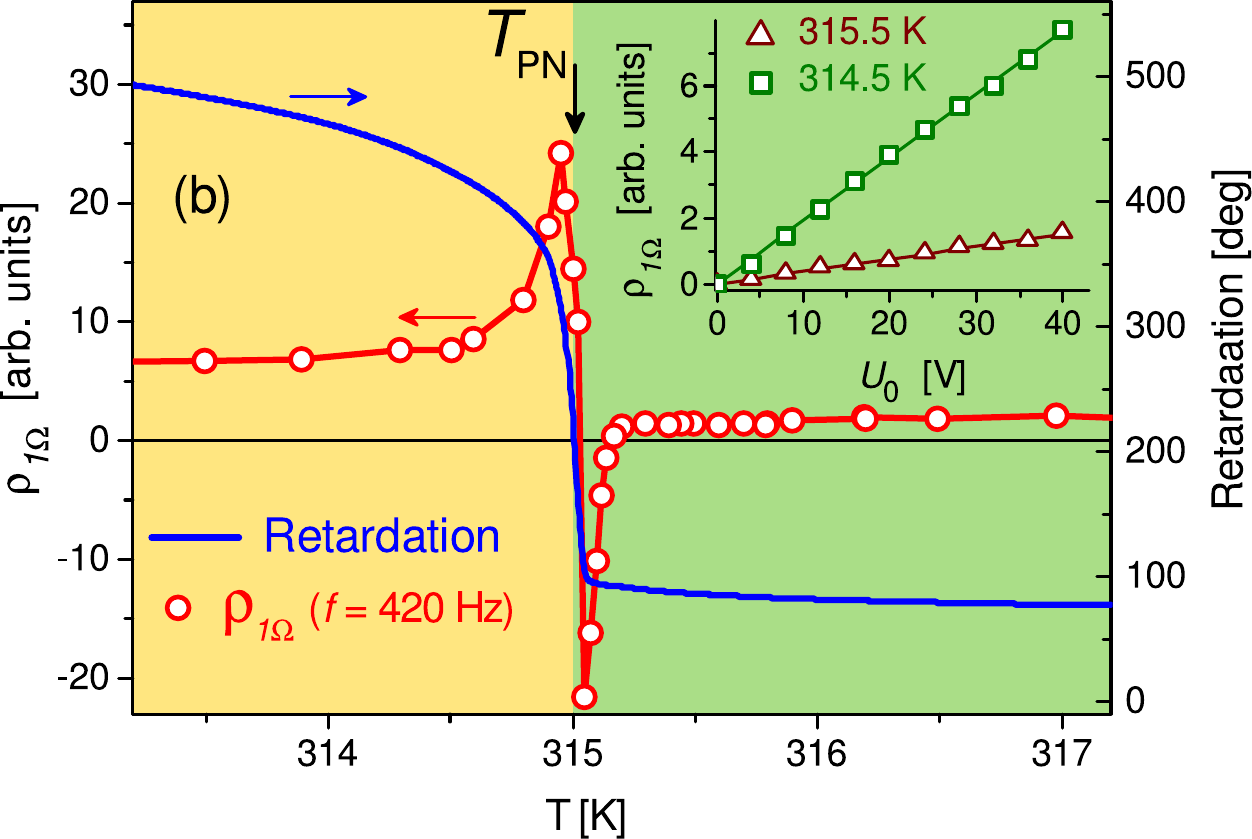}\hspace{5mm}
\includegraphics[width=0.95\columnwidth]{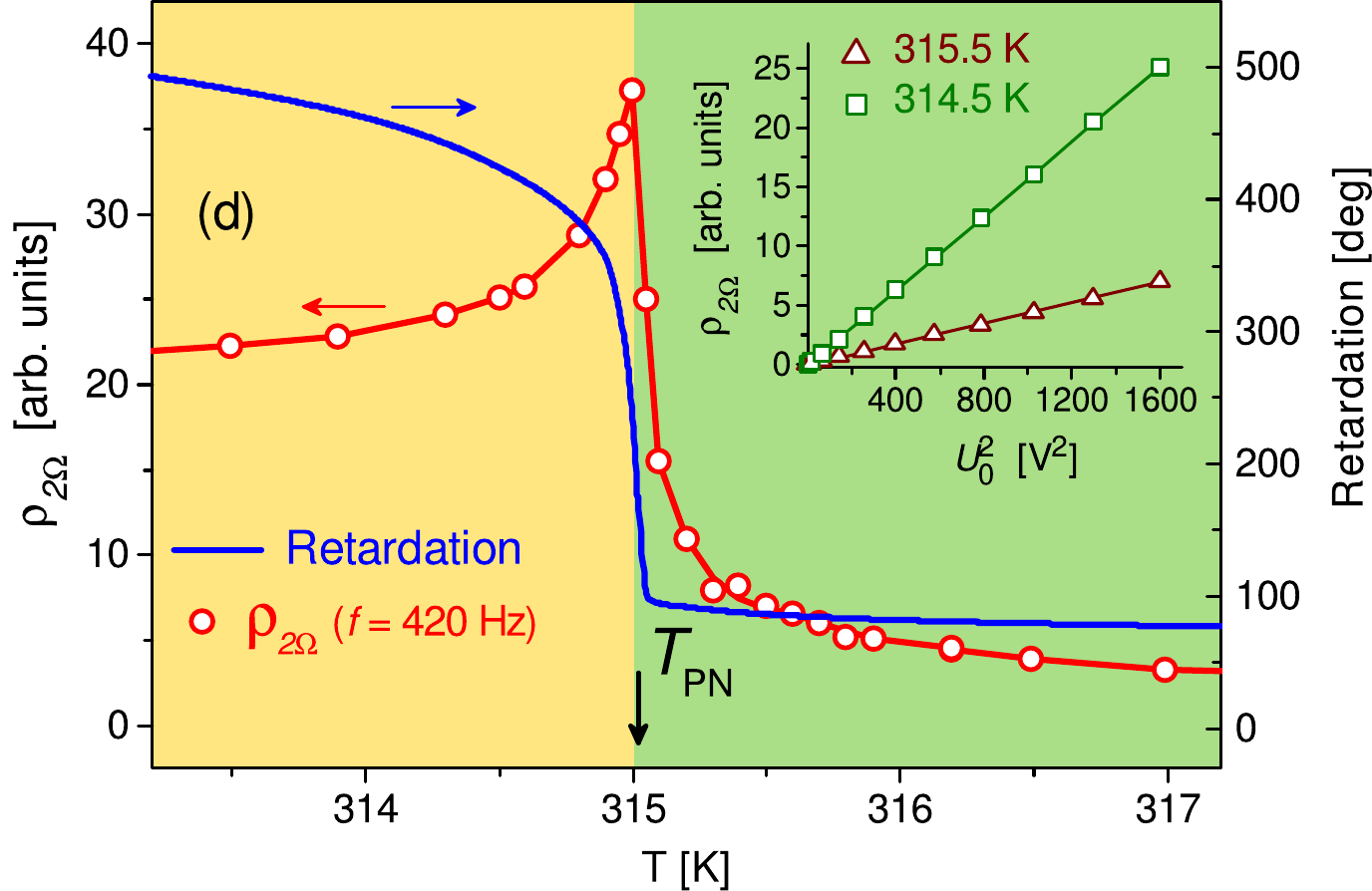}
\vspace{-0.0cm}
\caption{\textbf{Temperature dependence of the static nematic order in comparison to the resulting dynamic electro-optics of the AAO:7CB photonic metamaterial.} Temperature dependences of the amplitudes of (a,b) the first- and (c,d) second-harmonic electro-optic responses at a driving frequency of 420 Hz and a voltage amplitude of 40 V. (b,d) Electro-optic response in the vicinity of the paranematic-to-nematic transition. The temperature dependence of the optical retardation, $R$ is presented for comparison. Insets in (b) and (d), voltage dependences of the amplitude of the first and second harmonic electro-optic response, respectively, measured at fixed temperatures in the paranematic and
nematic states. The yellow and green backgrounds indicate the nematic and paranematic temperature ranges, respectively.}
\label{fig:exp_electrooptics_temp}
\end{figure*}

How is this thermotropic orientational order connected with the electro-optic response to an alternating external electrical field $\vec{E}$, see Fig.~\ref{fig:DynElectroOptics_3DMetMat}. In Fig.~\ref{fig:exp_electrooptics_temp} we show the amplitudes of the first [$\rho_{_{1\Omega}}$] and second
[$\rho_{_{2\Omega}}$] harmonics of the electro-optic response vs. temperature $T$ at a fixed driving frequency of 420~Hz
and a voltage amplitude of 40~V. Whereas the observation of a finite quadratic Kerr amplitude is expected as a response of the confined LC, we observe also a linear Pockels effect, at first glance symmetry forbidden, since a non-centrosymmetric structure of the nanocomposite is not apparent. 
The existence of both distinct electro-optic effects is unambiguously corroborated by the voltage dependencies of the amplitudes of the first and second harmonic amplitude, $\rho_{_{1\Omega}}(U)$ and $\rho_{_{2\Omega}}(U^2)$. As expected they scale linearly and quadratically on the amplitude of the applied voltage and thus to the applied external electrical field, respectively. 

We trace the linear electro-optic response of the confined mesogens to a preferential unipolar interfacial ordering that breaks the macroscopic centrosymmetry of the bulk isotropic or nematic ordering. However, given the cylindrical pore wall geometry this would still result in centrosymmetric structures. Thus we suggest that the conicity of the channels, as dictated by the fabrication process \cite{Cencha2019NondestructiveImbibition}, results in a macroscopic non-centrosymmetric polarizability along the channel axis and thus causes the Pockels effect. It is interesting to mention that a similar impact of conicity on second harmonic generation has recently been reported for chromophores embedded in conical silica nanochannels \cite{Kityk2021SHG}.

\begin{figure*}[htbp]
\vspace{-0.0cm}
\centering
\includegraphics[width=0.9\columnwidth]{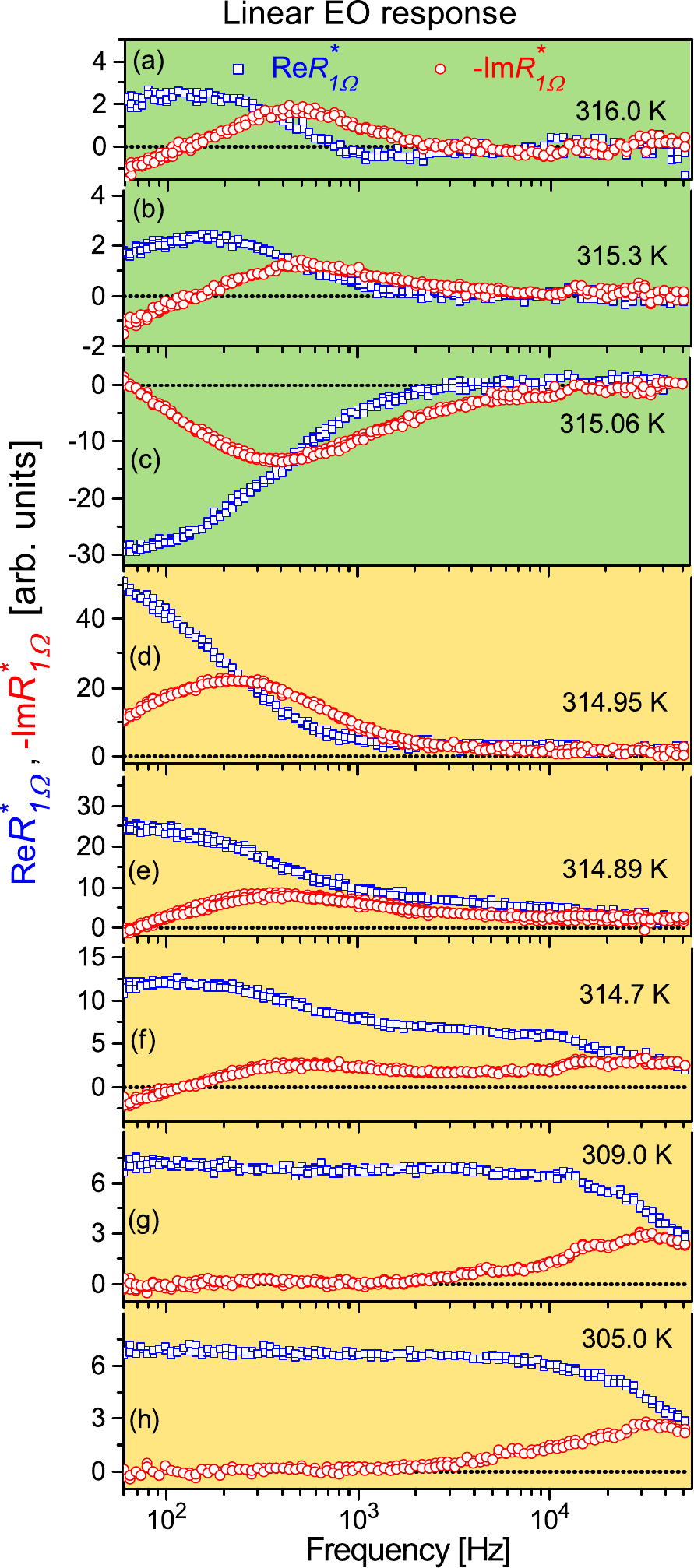}\hspace{10mm}
\includegraphics[width=0.9\columnwidth]{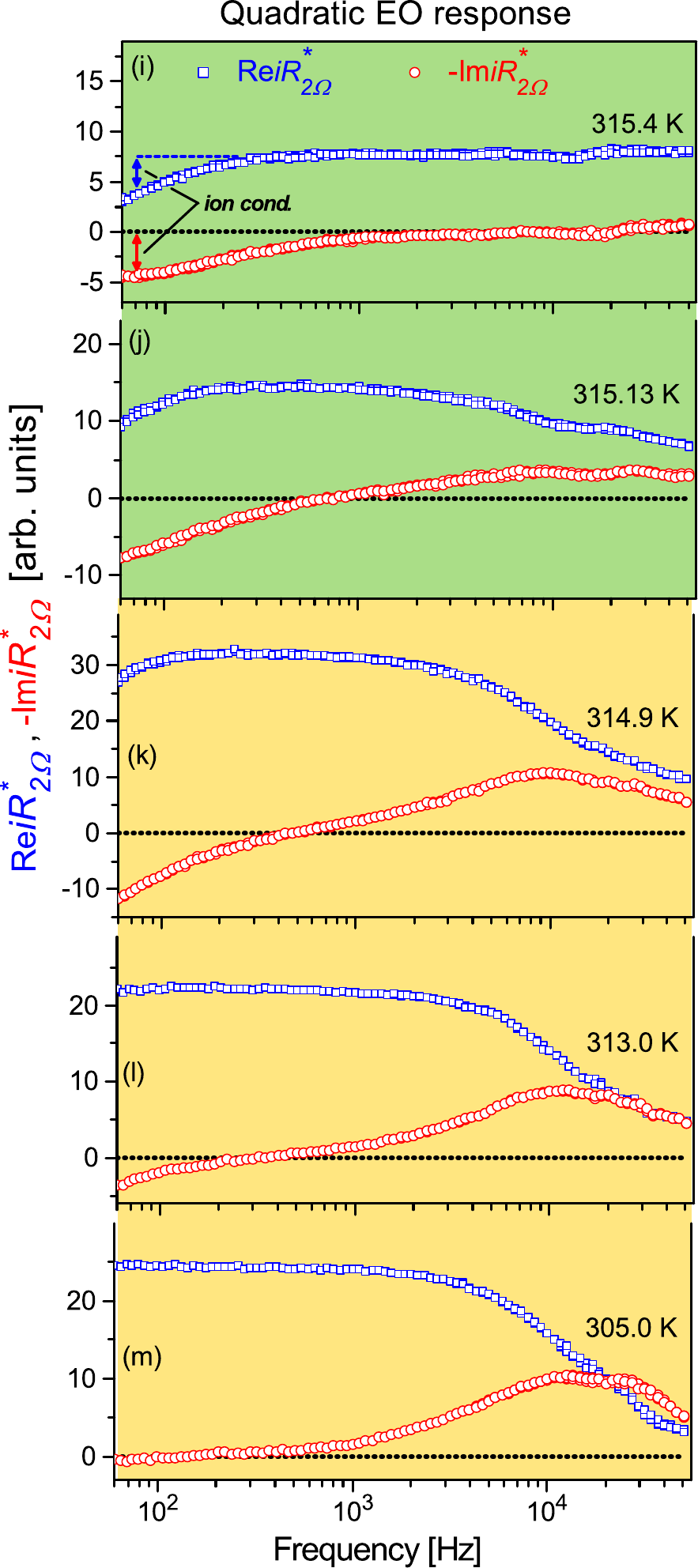}
\vspace{-0.0cm}
\caption{\textbf{Frequency-dependent electro-optics of an AAO:7CB photonic metamaterial above and below the paranematic-to-nematic phase transition}. Frequency dependence of the real and imaginary parts of (left) the first ($R^*_{1\Omega}=\rho_{_{1\Omega}}\exp(i\varphi_1)$, and  (right) the second
($iR^*_{2\Omega}=-\rho_{_{2\Omega}}\exp(i\varphi_2)$, right) harmonic electro-optic response measured at different temperatures in the confined paranematic (green background) and nematic states (yellow background).}
\label{fig:DynElectrooptics}
\end{figure*}

In Fig.~\ref{fig:DynElectrooptics} we show the frequency dispersion of the real and imaginary parts of the first ($R^*_{1\Omega}=\rho_{_{1\Omega}}\exp(i\varphi_1)$, left) and second
($iR^*_{2\Omega}=-\rho_{_{2\Omega}}\exp(i\varphi_2)$, right) harmonics of the electro-optic response measured at different temperatures. The dynamic electro-optic measurements are limited in our experiments to a frequency window of 0.05-50~kHz. Whereas its upper frequency limit is given by the instrumental constrains of the lock-in technique, the lower one originates from an ionic conductivity rising strongly at lower frequencies. It results in additional phase shifts and losses in the low frequency electro-optic responses below $\sim$200~Hz, especially at higher temperatures, see mark in Fig.~\ref{fig:DynElectrooptics}. The ionic conductivity influences both the linear and quadratic electro-optic responses and thus distorts the low-frequency dipolar dynamics. At driving frequencies above 400~Hz, however, the ionic conductivity is considerably suppressed. Therefore, we usually omit the frequency region below 400~Hz in our discussion.

In the paranematic phase $T>T_{PN}$ linear and quadratic electro-optic responses indicate evidently quite different frequency dispersions, compare the panels a) and i) in Fig.~\ref{fig:DynElectrooptics}. The linear response, symmetry forbidden in the bulk, exhibits a slow relaxation of Debye-like type observed in the frequency dispersions of its real ($\textnormal{Re} R^*_{1\Omega}$) and imaginary ($\textnormal{Im}R^*_{1\Omega}$) parts, see panel (a) and (b) in Fig.~\ref{fig:DynElectrooptics}. A characteristic relaxation maximum in the $\textnormal{Im}R^*_{1\Omega}(\omega)$-curve appears at a frequency $f_r=2\pi/\omega_r$ of about 0.5~kHz. 

The quadratic electro-optic response, by contrast, remains practically constant in the paranematic phase in the frequency range up to 50 kHz indicating thus a much faster relaxation with characteristic frequencies beyond the \emph{f}-window of the electro-optic measurements.

These distinct electro-optical dynamics originates in the heterogeneous molecular dynamics of molecular matter confined in nanopores. It has been demonstrated in a series of experimental studies that molecules show a slow molecular dynamics, in particular relaxation behavior near the LC-pore wall interface and a fast relaxation in the core region of the pore filling \cite{Aliev2005x,Aliev2010x,Calus2015ax,Calus2015Dielectric, Huber2015}. 

Thus the fast quadratic Kerr electro-optic response is dominated by the molecular dynamics originating from the isotropic (centrosymmetric) core region, whereas the linear Pockels effect ows his slow dynamics to the slow interfacial dipolar relaxations, in agreement with our conclusions above regarding its origin.

Cooperative molecular ordering in LCs, which accompanies phase transformations, may lead to an additional slowing down of the relaxation dynamics or even its anomalous behavior in the region of phase transformations \cite{Skarabot1999, Busch2017x}. Therefore, a peculiar behavior can be found for bulk liquid-crystalline materials. In the isotropic phase far above the clearing point nematic LCs (with dipolar mesogens) behave like dipolar liquids with a quadratic electro-optic response that diverges upon approaching the transition temperature to the nematic phase. According to the Landau-de Gennes free energy model for LC phase transitions \cite{deGennes1971,deGennes1969} the magnitude of the Kerr effect, as quantified by the Kerr constant, growths as $(T-T^*)^{-\gamma}$ where $T^*$ denotes the second-order pretransitional temperature (supercooling limit of the isotropic phase) and $\gamma$ is a critical exponent equal to 1 in mean-field approximation. This anomalous behavior was first observed by Tsvetkov and Ryumtsev \cite{Tsvetkov1968} and demonstrated in subsequent studies, see e.g. the Refs. \cite{Philip1992,Khoshsima2006,Schlick2018}.

Also here for the confined L-C state both the Pockels and the Kerr effect exhibit an anomalous behavior in the vicinity of the paranematic-to-nematic phase transition, see Figs.~\ref{fig:KKLZ_modeling} and ~\ref{fig:exp_electrooptics_temp}.

In the confined nematic phase, immediately below the transition temperature $T_{PN}$, the relaxation dynamics of the quadratic electro-optic response experiences a strong slowing down apparently caused by nematic ordering of the entire core region. The corresponding broad relaxation band in the dispersion curve $\textnormal{Im}iR^*_{2\Omega}(\omega)$ occurs in the frequency region of 10-20~kHz, see Fig.~\ref{fig:DynElectrooptics}(k-m). Interestingly, the slow relaxation of the linear electro-optic response, which is still dominated in close vicinity of $T_{PN}$ disappears at lower temperatures, see Fig.~\ref{fig:DynElectrooptics} (c-e). It is replaced by a considerably faster response dynamics with characteristic relaxation rates in the frequency range of 30-40 kHz as evidenced in panel g) and h) in Fig.~\ref{fig:DynElectrooptics}.

\begin{figure}[htbp]
\centering
\includegraphics[angle=90,width=0.9\columnwidth]{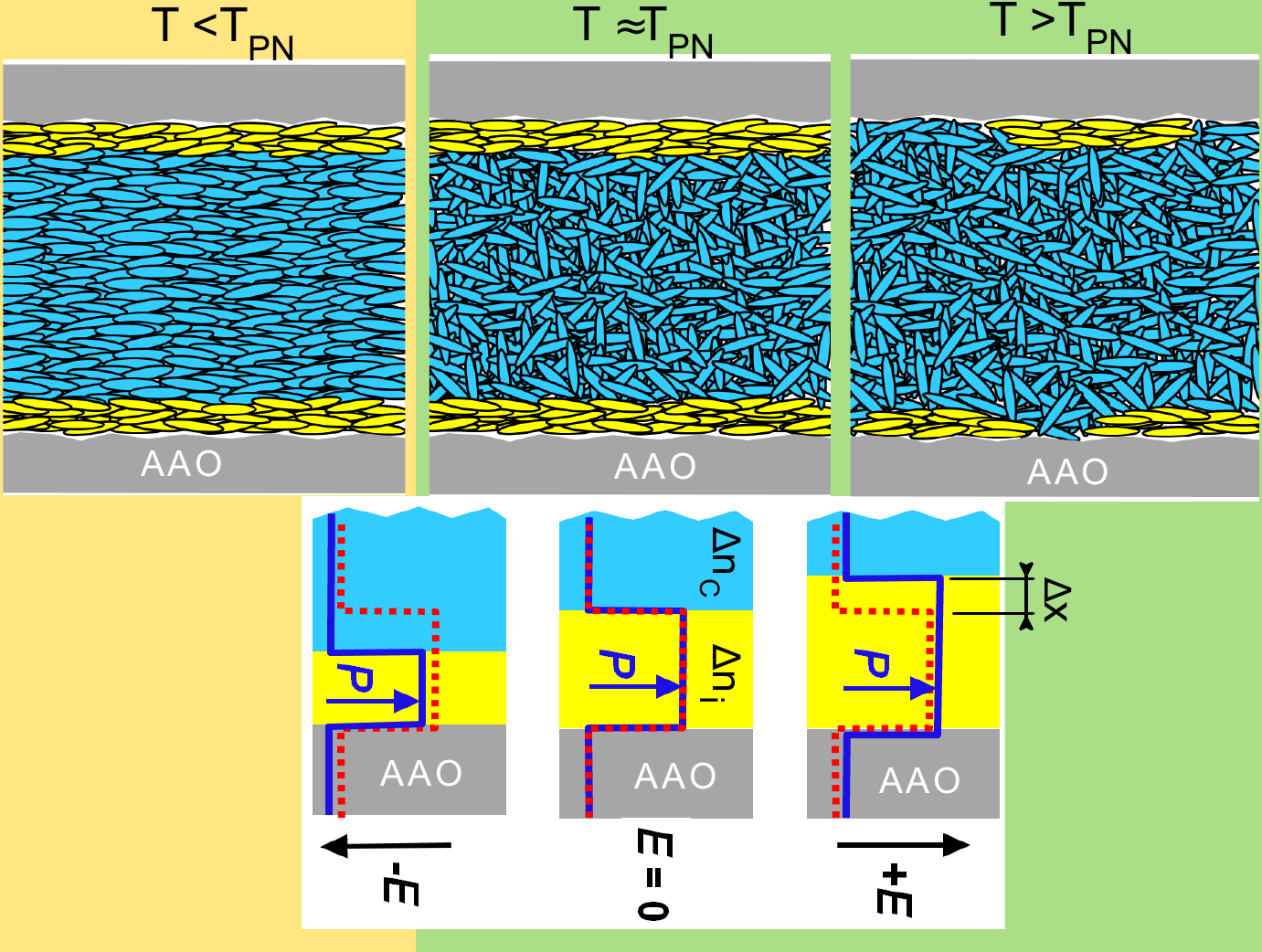}
\caption{\textbf{Illustration of temperature-dependent nematic order in the vicinity of the isotropic-to-nematic phase transition}. (left) Orientational ordering of the rod-like mesogens inside the cylindrical channel in the paranematic state ($T>T_{PN}$), in the vicinity of the phase
transition ($T \approx T_{PN}$) and in the nematic state ($T<T_{PN}$). (right) Sketch of the phase boundary (domain wall) separating the symmetry broken (non-centrosymmetric) interface region (yellow) and the centrosymmetric core region (cyan).}
\label{fig:confinedLC_illustration}
\end{figure}

The origin of this peculiarity deserves a more detailed discussion. According to Bengoechea and Aliev \cite{Aliev2005x,Bengoechea2005} and in agreement with subsequent dielectric studies by Calus \emph{et al.} \cite{Calus2015ax,Calus2015Dielectric} slow relaxations of confined liquid crystals suggest that the mobility of molecules in the surface layers is strongly affected by boundary conditions. Near-interfacial layers are characterized by a considerably larger
effective viscosity compared to the bulk one which solely slows down the dipolar relaxation and can thus not explain an abrupt change in the response dynamics. Therefore, we suggest that rather an  alternative mechanism based on interphase boundary dynamics in an alternating electric field is responsible for the peculiar dynamics.

The left sections of Fig.~\ref{fig:confinedLC_illustration} sketch the orientational ordering of the rod-like LC molecules inside a cylindrical channel in the paranematic state ($T > T_{PN}$), in the close vicinity of the phase transition ($T \approx T_{PN}$) and in the confined nematic state ($T<T_{PN}$), respectively. The right section sketches the corresponding phase boundary ("domain wall") separating the symmetry broken (non-centrosymmetric) interfacial polar regions and the centrosymmetric core region. An alternating electric field, applied along the pore axis, results in an oscillating radial shift of the phase boundary causing thereby an alternating optical birefringence, i.e., the dynamic electro-optic effect. 

Note that similar domain-like or soliton approaches have been successful in the explanation of anomalous behavior of dielectric and elastic properties in a number of ferroelectric and ferroelastic materials, see e.g. the  Refs.\cite{Kityk1996,Sondergeld2000,Schranz1999}. The corresponding domain wall dynamics is due to its collective character slow compared to the dynamics of individual molecules. This feature is evidently presented in the frequency dispersions of the imaginary part of the linear electro-optic response,
 $\textnormal{Im}R^*_{1\Omega}(\omega)$, recorded in the paranematic phase: A characteristic relaxation band appears in the frequency
 range of several hundred Hertz, see Fig. \ref{fig:DynElectrooptics} (a-d). The electro-optic contribution due to the alternating phase boundary dynamics is expected to be
 substantial whenever the birefringence of neighboring phase regions, i.e., the interfacial birefringence, $\Delta n_i$, and the core birefringence, $\Delta n_c$ substantially differ from each other, see  Fig.~\ref{fig:confinedLC_illustration}. 
 
 Analytically, an effective linear electro-optic coefficient
 $r \propto \Delta(\Delta n)/\Delta E \propto   (\Delta n_i-\Delta n_c)\cdot(\Delta x/\Delta E)$ results, where $\Delta x$ is the radial shift of the phase boundary (domain wall) induced by the applied electric field $\Delta E$. In the paranematic phase, $\Delta n_i\gg \Delta n_c=0$ thus the domain wall contribution with slow relaxation dynamics is evidently presented here, see panel a) and b) in Fig.~\ref{fig:DynElectrooptics}. Far above $T_{PN}$ the interfacial polar regions form just rare local islands. Their number grows however as the temperature decreases transforming them into closed surface layers upon approaching the paranematic-to-nematic transition temperature $T_{PN}$. This behavior is characterized by a pretransitional tail, i.e., a slowly rising retardation, evidently seen in the temperature dependence of the optical retardation above $T_{PN}$, see Fig. \ref{fig:exp_electrooptics_temp}.
 
 Note that in the close vicinity of $T_{PN}$ the core region is still isotropic ($\Delta n_c=0$) thus due to the large interphase area the linear electro-optic response is anomalous, see Fig.~\ref{fig:DynElectrooptics} (c, d) and Fig.~\ref{fig:exp_electrooptics_temp} (a, b). An abrupt almost step-like increase in the optical retardation right
 below $T_{PN}$ indicates a complete nematic ordering in the core region. 
 
 Considerably below $T_{PN}$ $\Delta n_i\approx \Delta n_c$ thus the domain like contribution, which is $\propto (\Delta n_i-\Delta n_c)$, becomes evidently suspended, see Fig.~\ref{fig:DynElectrooptics} (g, h). The remaining linear electro-optic response is fast and originates from the orientational dynamics of individual molecular dipoles in an alternating electric field. The surface induced polarity in the interfacial molecular layer may be transferred via intermolecular interactions towards the nematic core region of the pore filling. Accordingly, a larger number of molecular dipoles may contribute to the linear electro-optic response explaining its larger magnitude in the confined nematic phase compared to the paranematic one.

\subsection{Landau-De Gennes free energy analysis of confined liquid-crystalline order and Kerr electro-optics}
Whereas the description of the linear electro-optic response has qualitative character, the Kerr effect in the AAO:7CB nanocomposite and the quadratic electro-optic response may be described phenomenologically by employing the Kutnjak-Kralj-Lahajnar-Zumer (KKLZ)-model  \cite{Kutnjak2003,Kutnjak2004}. The orientational order inside the nanochannels
results in an excess optical retardation $\Delta R \propto S$ (see Fig.~\ref{fig:KKLZ_modeling}a, where $S=\frac{1}{2}\langle 3\cos^2\phi-1 \rangle$ is the scalar order parameter,
$\phi$ is the angle between the characteristic axis of the molecules and a direction of preferred local molecular orientation (the so-called director $\vec{n}$). 

The KKLZ model extends phenomenological the Landau-de Gennes approach towards spatially confined nematic phases by introducing a so-called nematic ordering field $\sigma$ which
couples bilinearly with the nematic order parameter \cite{Kutnjak2003,Kutnjak2004}. For the untreated host AAO matrix the channel walls enforce planar anchoring, whereas their elongated cylindrical geometry induces preferential ordering of guest rod-like molecules parallel to the long channel axes. The ordering field $\sigma$ is proportional
to the anchoring strength and inversely proportional to the pore diameter, i.e. it is defined by host-guest interactions at the channel wall interface and its geometrical
constraints. The KKLZ free energy is expressed in terms of the dimensionless scaled order parameter $q = S/S_0$, where $S_0 = S(T_{IN})$ stands for the nematic
order parameter value at the isotropic-to-nematic transition temperature of the bulk system, $T_{IN}$. The dimensionless free energy of the spatially confined nematic LC is then represented as follows:
\begin{subequations}
\begin{equation}
 f=f(q)+f(q,\sigma)+f(q,\kappa) + f(q,E),
 \tag{1}
 \end{equation}
\begin{align}
 f(q) = tq^2-2q^3 + q^4, \\
f(q,\sigma) =  - q\sigma, \\
f(q,\kappa) =   \kappa q^2, \\
f(q, E) = -a_1q E^2 -a_2 q^2 E^2.
\end{align}
\end{subequations}
where $t = (T - T^*)/(T_{IN}-T^*)$ is the dimensionless reduced temperature and $T^*$ is the effective temperature. Here $f(q)$  represents the free energy of
the bulk system, $f(q,\sigma)$ describes the couplings between the nematic order parameter $q$ and the geometric ordering field $\sigma$, $f(q,\kappa)$ accounts
for quenched disorder effects, $f(q,E)$ characterizes the coupling between the order parameter $q$ and an external electric field $E$ applied along the pore axis.

For symmetry reasons the electric field couples quadratically with the order parameter, i.e. two terms $q E^2$  and $q^2 E^2$ are presented in the Landau-de Gennes free
energy \cite{Rjumtsev1995}. The quenched disorder ($\kappa$-therm) results in a downward shift of the effective transition temperature $t_n$, whereas all other coupling terms rise their value as $t_n = 1+\sigma-\kappa + a_1 E^2 + a_2  E^2$. Here we assume that the electric coupling constants $a_1$ and $a_2$ are positive, i.e., an application of the electric field lowers the free energy of the axial nematic state making it energetically preferable. Moreover, the bilinear coupling term $q\sigma$,
likewise $q E^2$-term, induces a nematic ordering also above $t_n$. The geometrical constraint results in a precursor behavior with characteristic paranematic tails observed in the temperature dependence of the nematic order parameter. In the vicinity of the paranematic-to-nematic transition depending on a pore diameter the order parameter is characterized by a discontinuous subcritical ($\sigma < 1/2$) or a continuous overcritical  ($\sigma > 1/2$) evolution \cite{Kutnjak2003,Kutnjak2004, Kityk2008,Calus2012,Calus2014, Huber2020}.

Minimization of the free energy expansion (eq. 1) with respect to the order parameter $q$ at $E=0$ gives its equilibrium value $q_e$. It scales linearly with the measured excess retardation, $\Delta R$. A transfer from a reduced dimensionless temperature scale to the normal allows one to calibrate the model parameter $\Delta T^*=T_{IN} -T^*$ to 3.4 K, as it was derived in previous
polarimetric studies \cite{Calus2014}. A best fit in the region of the nematic-to-isotropic (paranematic) phase transition yields $\sigma=$0.18 and $\kappa=$0.15. Both values agree well with Ref.\cite{Calus2014} as extrapolations of $\sigma$- and $\kappa$-values obtained in a series of AAO:7CB nanocomposites with host matrices of smaller pore diameters. 

The solid line in Fig.~\ref{fig:exp_electrooptics_temp}a represents the best fit of experimental data points obtained within the KKLZ model. The dotted line traces the thermo-optic contribution (base line) obtained within the same fitting procedure. Overall, the KKLZ model describes remarkably well the behavior of the optical retardation in the paranematic state as well as in a broad temperature region of paranematic-to-nematic phase transition. Deviations far below $T_{PN}$ can be attributed to a saturation of the nematic ordering in the core region of pore filling that are originates the limitation to fourth-order terms  \cite{Calus2012}.
 
The amplitude of the quadratic electro-optic response, measured at an applied field $E=U/h$ ($U$ is the applied voltage), scales linearly with the difference $\Delta R(E)-\Delta R(0)$.
$\Delta R(E)$ may be derived similarly to $\Delta R(0)$ as described above, i.e., by the minimization of the free energy expansion (eq. 1) with respect to the order parameter $q$ taking into account the electrical coupling terms of the free energy defined by eq. 1c. The quadratic electro-optic response derived in such a way,
$\rho_{_{2\Omega}} = k[\Delta R(E)-\Delta R(0)] \propto [q_e(E) - q_e(0)]$ ($k$ is the scaling parameter) is shown in Fig.~\ref{fig:exp_electrooptics_temp}b.
The measured quadratic electro-optic response is presented for comparison. Here the coupling constants
$a_1$=1.0$\cdot$10$^{-13}$ cm$^2$/V$^2$ and  $a_2$=1.3$\cdot$10$^{-12}$ cm$^2$/V$^2$ have been chosen in a way that provides the values of field induced
optical retardation close to those ones observed in the experiment. 

The presented phenomenological model well reproduces basic features of the anomalous electro-optic behavior, particularly the asymmetric peak of the quadratic electro-optic response. A discrepancy between the theory and experiment, evidently seen in the confined nematic state, may have several reasons. Our phenomenological approach considers the static electro-optic response ($\omega=$~0), whereas the experiments explore the quasi-static electro-optic response ($\omega/ 2 \pi =~$420~Hz). The chosen frequency can be considered as a compromise providing, on the one hand, a reduced influence of the ionic conductivity, and on the other hand, still insignificant dipolar relaxation losses. However, at least a weak influence of the ionic conductivity on the measured electro-optic response, which is considerably temperature-dependent, indeed is presented. Also spatial inhomogeneities of the channel diameters (up to 10\% radius variation) can result in gradient field effects and likewise field depolarization at channel interfaces.

\section{Conclusion}

We presented a study on the electro-optical dynamics of the rod-like liquid crystal 7CB embedded into parallel arrays of cylindrical nanopores in anodic aluminum oxide. Linear and quadratic electro-optical effects are explored via first- and second-harmonic polarimetry responses that are measured by a lock-in technique in a frequency window from 0.05 - 50 kHz and a broad temperature range comprising a phase transition between a confined paranematic and nematic state.

We find that a linear electro-optical response, symmetry forbidden in bulk nematics, occurs in both states due to polar ordering at the nanopore walls. The dynamics of the quadratic electro-optical response reveals anomalous dynamical behavior at the paranematic-to-nematic phase transformation that can be quantitatively described by a Landau-De Gennes free energy model. The linear and quadratic electro-optical effects are characterized by substantially different molecular relaxation rates caused apparently by a slow and fast mesogen reorientation in the proximity of the pore wall and in the pore center, respectively. 

One can envision that the observed integrated electro-optics that sensitively depends on multiphyiscal couplings is particularly interesting to induce lateral phase shift gradients for electromagnetic waves in solid supports or surfaces at will by external electrical field and temperature gradients. Such active photonic metamaterials are demanded for the design of in-operando adjustable optical components \cite{Yu2011, Pendry2012, Buchnev2015, Nemati2018, ZhangM2018, Shaltout2019}.

Thus, similarly as it has been demonstrated for liquid-infused porous structures with electrochemical actuation \cite{Brinker2020GiantMaterial} and wetting changes \cite{Xue2014} as well as versatile multifunctional material behavior in general \cite{Style2021SolidLiquidMaterials} our study indicates that liquid crystal-infused nanoporous solids allow for a quite simple fabrication of electro-active functional nanomaterials. The simple functionalization approach by capillarity-driven spontaneous imbibition of the liquid-crystalline melt in combination with tailorability of anodic aluminum oxide \cite{Lee2014, Chen2015, Sukarno2017, Busch2017} and the availability of other self-organized, optically transparent mesoporous media \cite{Gallego2011, Sousa2014, Huber2015, Spengler2018} indicates also the versatile character of the presented material fabrication process.

Finally, our study is a fine example, how the combination of soft and hard matter opens up the possibility of using multi-scale self-assembly and phase transitions to transport peculiar multi-physical couplings in geometrical confinement from the nano- via the meso- to the macroscale in order to design robust hybrid materials with integrated functionality, similar as found in many biological composites \cite{Eder2018, Gang2020SoftMatterBook}.

\section*{Author Contributions}
A.V.K., A.A., M.R., N.A., and P.H. conceived and designed the electro-optic experiments. A.V.K, M.L., and P.P. fabricated the nanocomposites. A.V.K and M.N. performed the optical experiments. A.V.K. and Y.S. analyzed the data. A.V.K., Y.S., and P.H. interpreted the experiments. A.V.K. and P.H. wrote the manuscript. All authors proofread and edited the manuscript.

\section*{Acknowledgements}
We received funding from the European Union's Horizon 2020 research and innovation programme under the Marie Skodowska-Curie grant agreement No. 778156.
Support from resources for science in the years 2018-2022 granted for the realization of the international co-financed project No. W13/H2020/2018 (Dec. MNiSW 3871/H2020/2018/2) is also acknowledged. We also profited from support by the Deutsche Forschungsgemeinschaft (DFG, German Research Foundation) within the collaborative reserach center SFB 986 ''Tailor-Made Multi-Scale Materials Systems'', project number 192346071).
\vspace{0.5cm}



\balance



\bibliographystyle{rsc}
\providecommand*{\mcitethebibliography}{\thebibliography}
\csname @ifundefined\endcsname{endmcitethebibliography}
{\let\endmcitethebibliography\endthebibliography}{}

\end{document}